# Supershape nanoparticle plasmons

F. Babaei[1, *], M. Javidnasab[2], A. Rezaei[3]

[1] Department of Physics, University of Qom, Qom, Iran
[2] Faculty of Physics, University of Tabriz, Tabriz, Iran
[3] Faculty of Chemical and Petroleum Engineering, University of Tabriz, Tabriz, Iran
*) Email: fbabaei@qom.ac.ir

## Abstract

In this work, we proposed new shape nanoparticles in the name of supershape nanoparticles by manipulation in the morphology of a disk nanoparticle. The electric field distribution of supershape nanoparticles were investigated at resonance wavelength of particle plasmons. The effects of dispersed medium and different sensory materials on particle plasmons were reported. The obtained results showed that there exist the multiple plasmonics modes in supershape nanoparticles. We found that the high sensitivity factor is available in the supershape nanoparticles. This study can be a base for the characterization of multiple particle plasmons in plasmonic devices for sensing applications.

Keywords: plasmons; nanoparticles; sensing

## I. Introduction

Fabrication of nanoparticles in an arbitrary and pre-controlled shape can be more interesting and attractive. Based on our knowledge, there is not still a global physical technique with a few steps of fabrication. Although, one can use the lithography method to implement especial patterns on the surface of substrates which is, of course, several steps. On the other hand, the oblique angle deposition method has been proposed for engineering the columnar morphologies along their growth direction, from matchsticks and zigzags to continuously curved shapes and helices in nano dimensions [1-3]. Also, the growth of photonic crystals has been down with combination of oblique angle deposition and lithography [4-6]. But, a routine method does not exist for fabrication of nanoparticles with complex morphologies. The technology of print in nanoscale can helps to people [7-9]. Maybe in near future, by designing nano-scale templates with printing technology and the use of ink of nanoparticles, any kind of complex structure can be fabricated.

Plasmonics plays an essential role in detection of adsorbed nanoparticles on a surface [10], bacterias [11], proteins [12], DNA [13], RNA [14], viruses [15], analytes [16], chemical and biological species [17]. The plasmonic sensors based on metal nanoparticles and metal nanostructures especially including Silver and Gold. The resonant interaction between electron-charged oscillations near the surface of the metal and the electromagnetic field of the light creates the localized plasmons [18]. It is well known that the characteristic of localized plasmons in nanoparticles strongly depends on the size, shape, the interparticle separation, the refractive index of the sensed medium, and dispersed material [19]. In most cases, the sensory elements are not spherical, but they are non-spherical and more complex [20]. For detection a complex nanostructure, the surface to volume ratio of the sensor must be high. Mapping localized plasmons on single metal nanoparticle and dimers has been suggested [21,22]. Also, for excitation of surface plasmon-polaritons at planar interface of a metal thin film and a dielectric medium, the different configurations such as Otto, Kretschmann and Sarid have been designed [23,24]. In recent years, the surface multiplasmonics has been introduced using a homogeneous metal thin film and a nonhomogeneous dielectric sculptured thin film so that it can be used in sensors for simultaneous detection of more than one chemical species [25-27].

A group of nanoparticles can be designed that the closed boundary of them follows from the Gielis superformula [28]. The Gielis superformula was used in plasmonics for describe the shape of plasmonic nanoparticles in 3D [29] and for the inverse design of translational-invariant plasmonic cylinders in 2D [30,31]. Also, a strategy has been proposed to design the morphology of metal nanoparticles, maximizing the electric field average on their surface [32]. The Gielis superformula in two dimensions has the mathematical expression in polar coordinates:

$$r(\varphi) = \left( \left| \frac{\cos(\frac{m_1 \varphi}{4})}{a} \right|^{n_2} + \left| \frac{\sin(\frac{m_2 \varphi}{4})}{b} \right|^{n_3} \right)^{-\frac{1}{n_1}}$$

(1),

where $r$ and $\varphi$ are respectively the radius and polar angle. The different shapes can be generated by selecting of the different values for the parameters a, b, $m_1$, $m_2$, $n_1$, $n_2$ and $n_3$.

Here, we used the Gielis superformula to generation of a kind of nanoparticles and called them as the supershape nanoparticles (SNPs). In our work, gold metal is selected for simulation. The optical extinction, scattering, absorption and electric field distribution of SNPs were investigated. The influence of the refractive index of the dispersed material and sensed medium in SNPs is studied. In addition, the sensitivity factor (SF) of all the plasmonic modes in SNPs is reported. Method of simulation is given in Section II and followed by results and discussion in Section III.

## II. Method of simulation

We used 3D finite difference time domain (FDTD) method for simulations, that is one of the most used methods for solving Maxwell's time-dependent equations, especially when the analytical solutions are extremely difficult to obtain [33]. The optical constant of gold is taken from the Johnson and Christy experimental data [34]. In this work, we set the mesh steps 0.5 nm for all of the three directions ($\Delta x = \Delta y = \Delta z = 0.5\ nm$) to get convergent results. The perfectly matched layer (PML) boundary condition is used for the boundaries of the simulation.

### III. Results and discussion

We considered SNPs as the closed boundary of them in xy-plane obtained from the Gielis superformula. In all calculations, the parameters of $m_1 = m_2 = m$ and $a = b$ were fixed in Eq.1 and also four variables as (m $n_1$ $n_2$ $n_3$) describe the morphology of SNPs. These geometries naturally exist in plants [35]. Herein, three kinds of SNPs are assumed to names of curved NPs (Fig.1a1-1a6), pod NPs (Fig.1b1-1b6) and starfish NPs((Fig.1c1-1c6)). This category was done based on the boundary of nanoparticle and it is quite arbitrary and optional. The average size in xy-plane and the thickness of SNPs were assumed 56 nm and 5 nm, respectively. In order of simulation, the center of coordinate system was in the center of SNP. In all plots, the initial electric field ($\vec{E}$) is polarized in the x-direction and propagates ($\vec{K}$) in the -z-direction (in normal direction to height of super-NP) (see Fig.1).

Electric field distributions in xy-plane for curved NPs, pod NPs and starfish NPs dispersed in air medium are respectively depicted in Figs.2,3&4 at maximum wavelength where it occurs in extinction spectra. It is observed in Fig.2a1 that hot spots are formed on sides A and A′ due to oscillation of the conductive electrons in the x- axis direction. In Fig.2a2 clearly obvious that enhancement of electric filed occurs in the vertices of A and A′. Because on outside the surface of near the sharp points or edges, the surface charge density has high values. In Fig.2a3, the major axes of ellipse have located in the y-axis direction and thus the electric field can reach to high values on sides A and A′. It is seen Fig.2a4 that the hottest spot happens on the bow A because more length of this bow is exposed to applied electric field and also there are two hot spots on the bows B and C with low density. In Fig.2a5, the hot spots have occurred on the bows A, B and in front of them (A′, B′) caused by the same reasons previously mentioned. In Fig.2a6, we see the pattern of electric field distribution is close to Fig.2a1 because the shape of NP is almost like a circle. For pentapod NP in Fig.3b1 hot spots are formed in the vertices of A, B, C due to the location of them on or near the x-axis. The position of hexapod NP in Fig.3b2 is such that the vertex A, front of it and other vertices are respectively located in the direction of and near to normal to applied electric filed. Thus, the hot spots have occurred on the vertices in the direction of x-axis and other vertices less affected. In Figs.3b3 and 3b4, there are two heptapod NPs that the former and latter include the sharper and wider vertices, respectively. The points A and B (see Fig.1) are located on y-axis then the surface charge density is not strong on them but other vertices depending on whether they are on or off the x- axis can contribute to the formation of hot spots. Also, there exist two octapod NPs in Figs.3b5 and 3b6 that the hottest spots can be formed on vertices A, B and front of them due to the above mentioned reasons although it is seen that hot spots have been created on the other vertices. The surface charge density is high in the former because of the sharpness of the vertices. We considered a starfish NP with three arms in Fig.4c1 that the azimuthal angle ($\varphi$) between arms was 120°. The hottest spot can be formed on vertex A and also there are hot spots on vertices B,C that the azimuthal angle of them is 30° with y-axis. These spots can be related to interaction photons and quantum of electronic surface charge density oscillations in the name of particle plasmons (localized plasmons). In Fig.4c2, the arms of starfish NP have an angle $\varphi=45°$ with x-axis, thus aggregation of the surface charge density creates the hot spots on the vertices A, B and front of them. In Fig.4c3, two arms of starfish are located in direction of normal to applied electric filed, then the hot spots cannot be formed on the vertices of them. But other arms (A, B and front of them) had an azimuthal angle 30° with x-axis that it causes the hot spots occur on surface of them. In Fig.4c4, only the points A, B, C (see Fig.1) have been located in direction of or near to applied electric field and this matter leads to creation of hot spots on them. Also in Figs.4b5 and 4b6, the arcs A and front of it are exposed to applied electric field thus the surface charge density can reach to high values in these parts of starfish NP.

The simulated extinction, absorption and scattering spectra for heptapod NP in the dispersed and sensed mediums as a function of wavelength are depicted in Fig.5.These mediums were chosen from materials with refractive indexes 1.0,1.33,1.36,1.46 and 1.49 respectively for air, water, ethanol, carbon tetrachloride and toluene. Herein, we considered the dispersed medium that means the whole volume of the nanoparticle is immersed in it and the sensed medium there was a difference that the substrate of nanoparticle was fixed as glass with refractive index 1.52. It is observed that the scattering of nanoparticles is negligible and then the extinction and absorption spectra have almost close values. It is well known that the optical scattering decreases when the size of the nanoparticle increases and vice versa. Then, we can ignore from the scattering of nanoparticles in spectra. We found two and four plasmonic modes respectively for heptapod NP in the dispersed and sensed mediums. These modes were extracted from extinction spectra where the resonances of particles plasmons have occurred in them. The first and second plasmonic modes (M1, M2) were, respectively, located at wavelengths (427,659) nm, (427,788) nm, (427,803) nm, (427,845) nm and (427,860) nm for heptapod NP dispersed in air, water, ethanol, carbon tetrachloride and toluene. The peak located at a short wavelength, is called the transverse plasmon mode (TM), while the other located at a longer wavelength, is called the longitudinal plasmon mode (LM). It is clearly seen that in inset plots as the refractive index of dispersed medium increases, the TM and LM were fixed and linearly rises, respectively.TM and LM respectively can be related to formation quadrupole and dipole in nanoparticle. When heptapod NP is deposited on glass substrate there exist four plasmonic modes for different sensory materials except air (see inset plots in Fig.5). The first TM mode (M1) was fixed at 427 nm and the second TM mode(M2) was located at 551 nm. The two LM modes (M3, M4) located at wavelengths (664,772) nm, (680,829) nm, (680,840) nm, (690,860) nm and (690,870) nm for air, water, ethanol, carbon tetrachloride and toluene, respectively. The existence multiple TM and LM may be caused by anisotropy dielectric permittivity in an environment nanoparticle or created by nanoparticle due to several pods. The calculations were also performed for other SNPs and the results of them have not been shown here because the obtained results for those cases were similar to Fig.5.

In order to further analyze for sensing applications, the sensitivity factor(SF) of all plasmonic modes for SNPs in dispersed and sensed mediums (SFd and SFs) are drawn as clustered column charts in Fig.6. The SF can be determined from slope plots the wavelengths of localized plasmons as a function of refractive index as $\Delta \lambda_{LP} / \Delta n$ (nm/RIU) that they are not shown here (see for a sample in inset plots for heptapod NP in Fig.5). A comparison between the SF values of curved NPs (a1-a7) showed that the first mode in both cases (SFd and SFs) is not sensitive to the refractive index and it remains constant by changing the refractive index. It is valid for pod NPs and starfish NPs. We found that the most value of SF belonged to cubic shape (see a5 geometry in Fig.1) in curve NPs so that it was $SFd_{M3}$=483.6 nm/RIU in third mode and $SFs_{M4}$=254.6 nm/RIU in forth mode. Also, for this cubic shape there exist a negative sensitivity about $SFs_{M3}$= -14.0 nm/RIU in third mode when NP is deposited on glass substrate. Here, the negative sensitivity means that the wavelength of particle plasmon decreases with increasing of refractive of sensed medium. It is obvious that for all SNPs the values of SFd are bigger than SFs due to coverage of the bottom of NP in sensed medium by glass. The obtained results revealed that when the more partitions of NP be elongated in the direction of applied electric field, the SF value can be increased (see a1 and a2 in Fig.1). The maximum value of SF in pod NPs was observed in third mode at SFd= 493.1 nm/RIU and SFs=483.0 nm/RIU, respectively in b1 and b5 shapes. It is seen that often the number of plasmonic modes in sensed medium is more than dispersed medium. Because the wavevector some of hidden plasmonic modes is greater than photon wavevector thus there is not an enough momentum for excitation of particle plasmons in surface of NP and the glass substrate can compensate this momentum. The obtained results for starfish NPs showed the maximum values of SFd and SFs were for c3 and c2 with $SFd_{M4}$=847.9 nm/RIU and $SFs_{M3}$=460.9 nm/RIU, respectively.

Finally, we moved from a disk nanoparticle(a1) to sun nanoparticle(c6). In this way, the morphology of nanoparticle is manipulated by creation of vertices, pods and arms. Our simulations reveal that the high sensitivity is available if plasmonic devices to be fabricated on basis of SNPs. Also, the existence of multiple plasmonic modes helps to sensing more than one adsorbed nanoparticles on the surface of SNPs.

## IV.     Conclusions

In this research, we theoretically studied the plasmonic properties of gold supershape nanoparticles such as curved nanoparticles, pod nanoparticles and starfish nanoparticles. The electric field distributions of supershape nanoparticles at maximum wavelength of particle plasmons were investigated. The resonance wavelength of particles plasmons is extracted from extinction spectra. The effect of refractive index on particles plasmons was considered by chosen the different sensory materials. The obtained results showed that the electric filed is enhanced in the parts of supershape nanoparticles when they are located in direction of applied electric field. It is revealed that for supershape nanoparticles on glass substrate there exist more plasmonic modes. The existence multiple plasmonic modes may be caused by anisotropy dielectric permittivity in an environment nanoparticle or created by nanoparticle due to several vertices, pods and arms. We found that the high sensitivity factors were 483.6 nm/RIU in third mode for cubic shape in curved nanoparticles, 493.1 nm/RIU in third mode for pentapod in pod nanoparticles and 847.9 nm/RIU in fourth mode for starfish with six arms in starfish nanoparticles. The details of this study can be applied in plasmonic sensors for simultaneous detection of more than one chemical species.

**Acknowledgments**

This work was carried out with the support of the University of Qom and University of Tabriz.

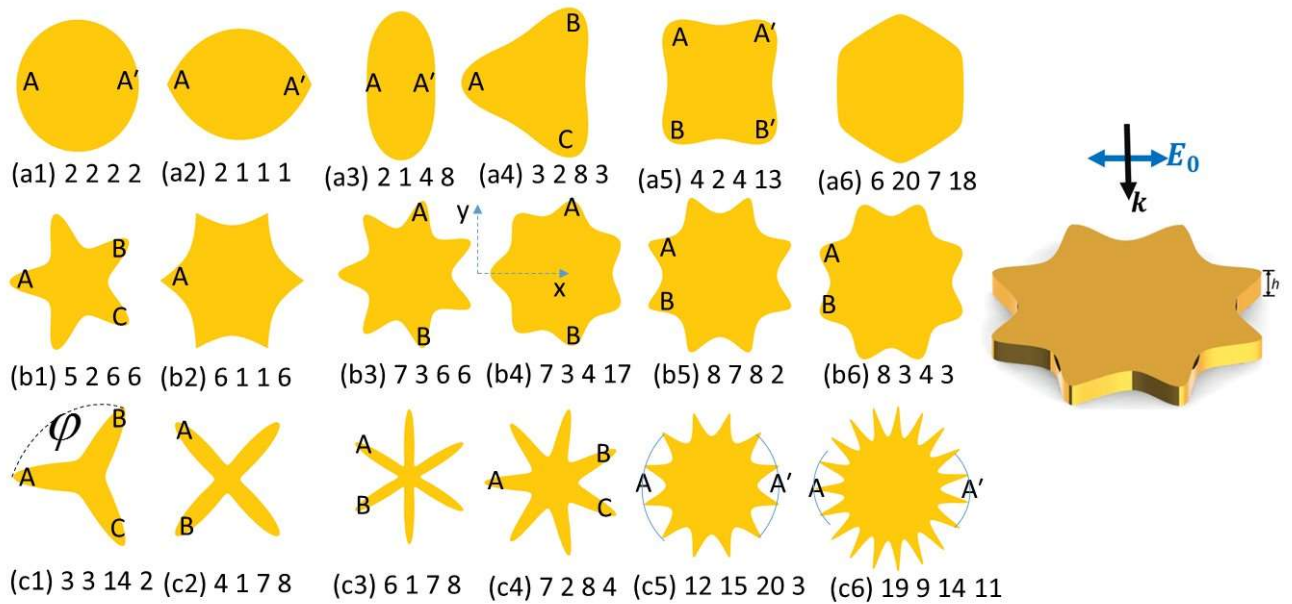

**Fig.1.** A schematic of SNPs. The closed boundary of NPs in xy-plane follows from Gielis superformula [ 28]. Four indexes (m $n_1$ $n_2$ $n_3$) indicate to morphology of NPs, (a1-a6) curved NPs, (b1-b6) pod NPs and (c1-c6) starfish NPs. The $\vec{K}$ and $\vec{E}$ directions for a sample are specified .

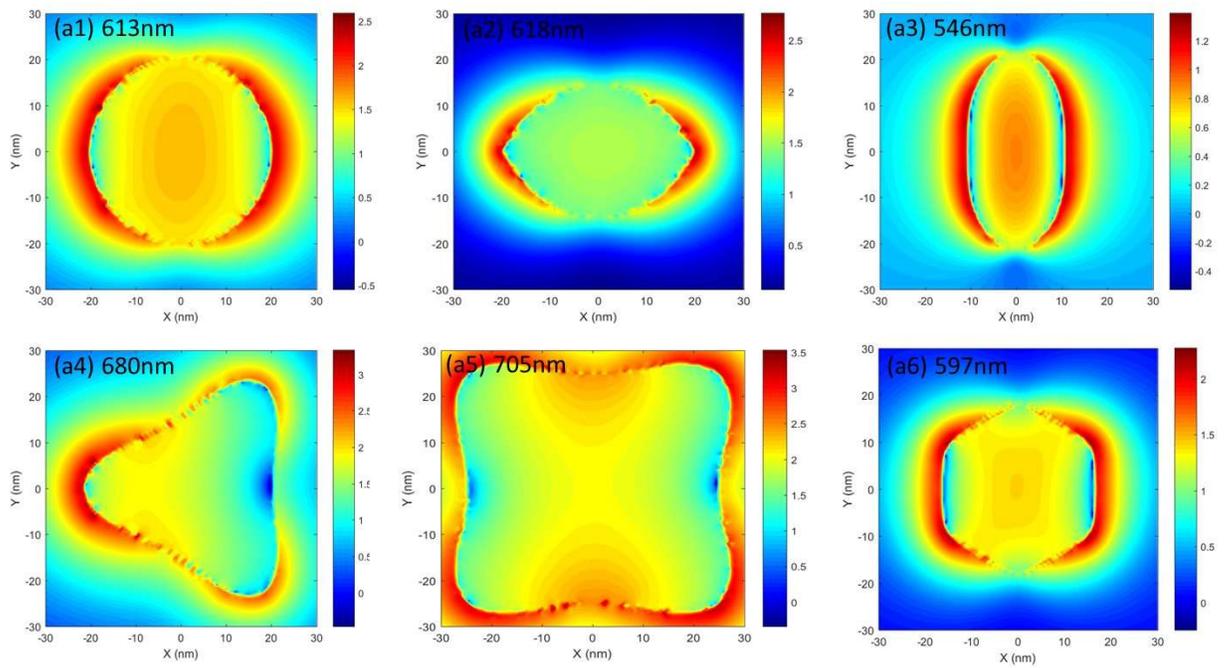

**Fig.2.** Simulated the electric field enhancement ($Log|\vec{E}|^2$ ) in xy-plane at maximum wavelength for curved NPs dispersed in air medium.

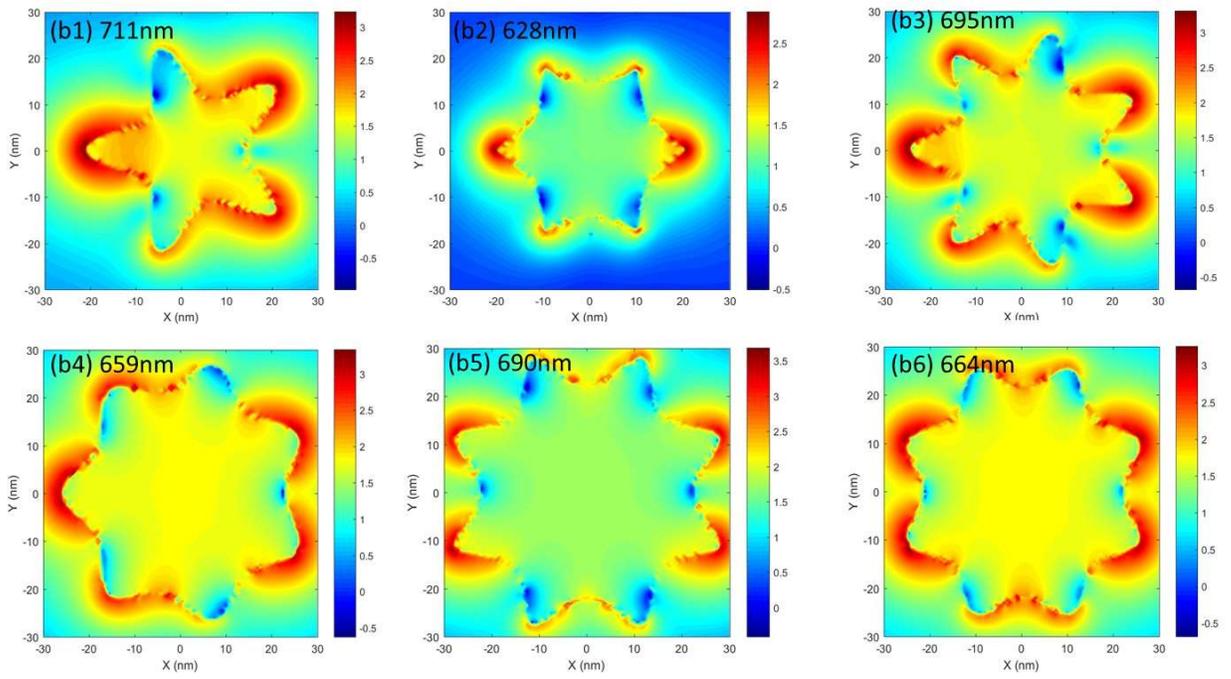

**Fig.3.** Same as Fig. 2, except that it is depicted for pod NPs.

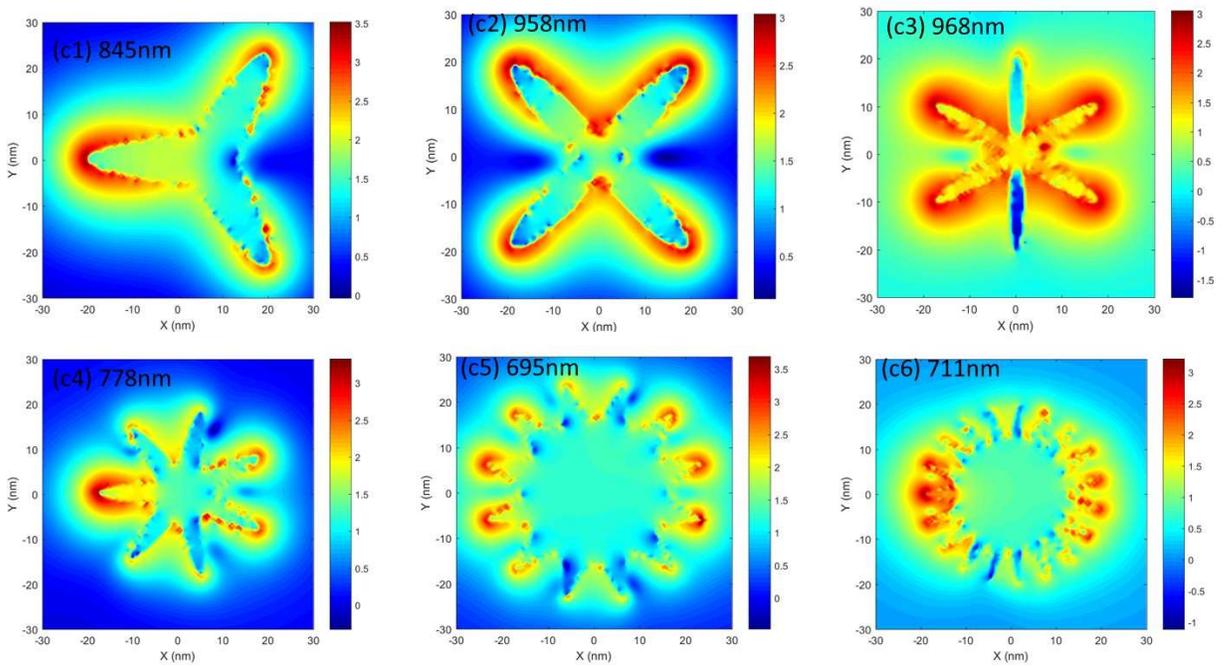

**Fig.4.** Same as Fig. 2, except that it is depicted for starfish NPs.

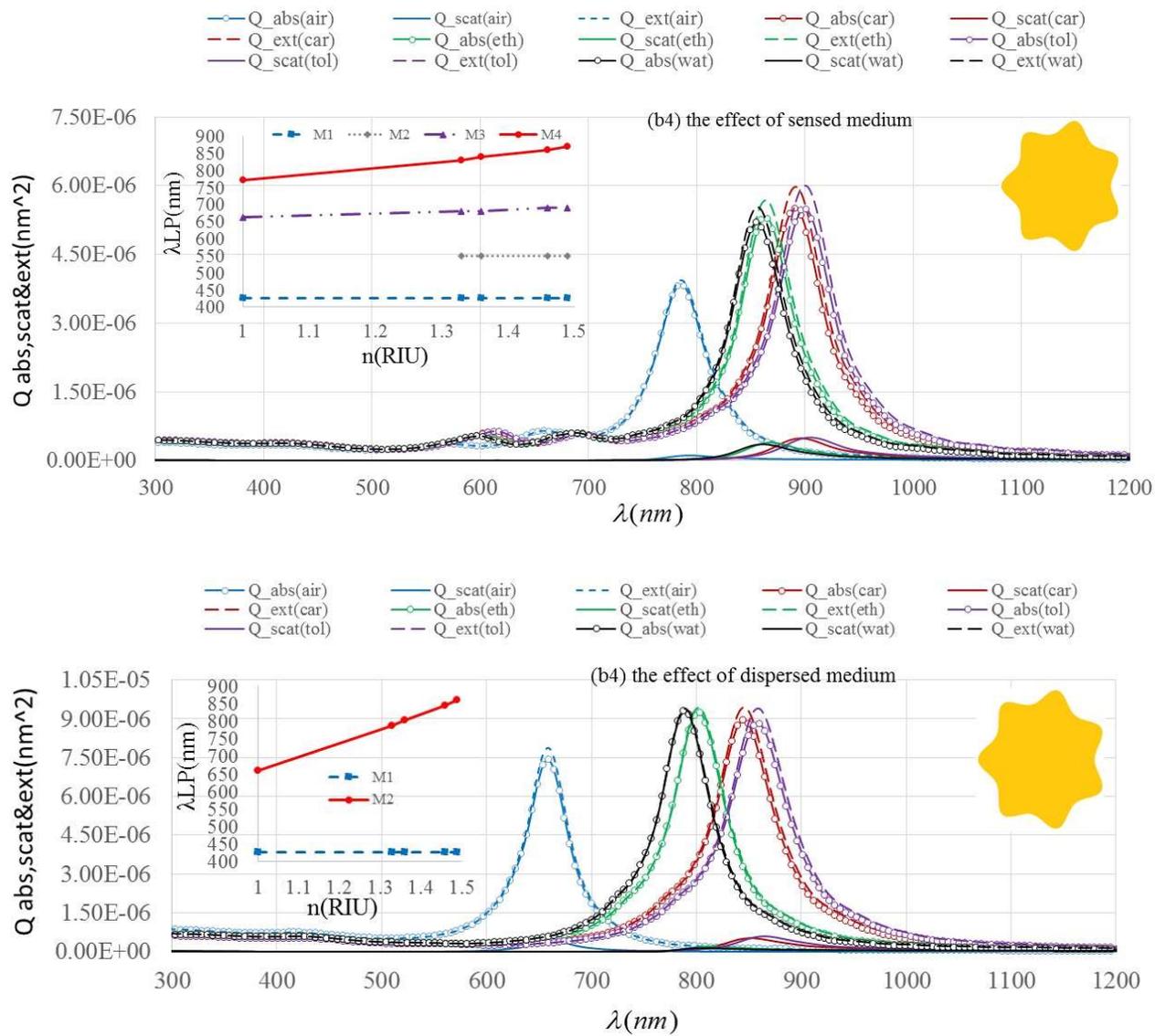

**Fig.5.** Calculated optical extinction, absorption, and scattering for heptapod NP (see b4 in Fig.1). In bottom and top rows, the effects of dispersed and sensed mediums are shown, respectively. Inset plots the wavelengths of particle plasmons (localized plasmons) extract from extinction spectra.

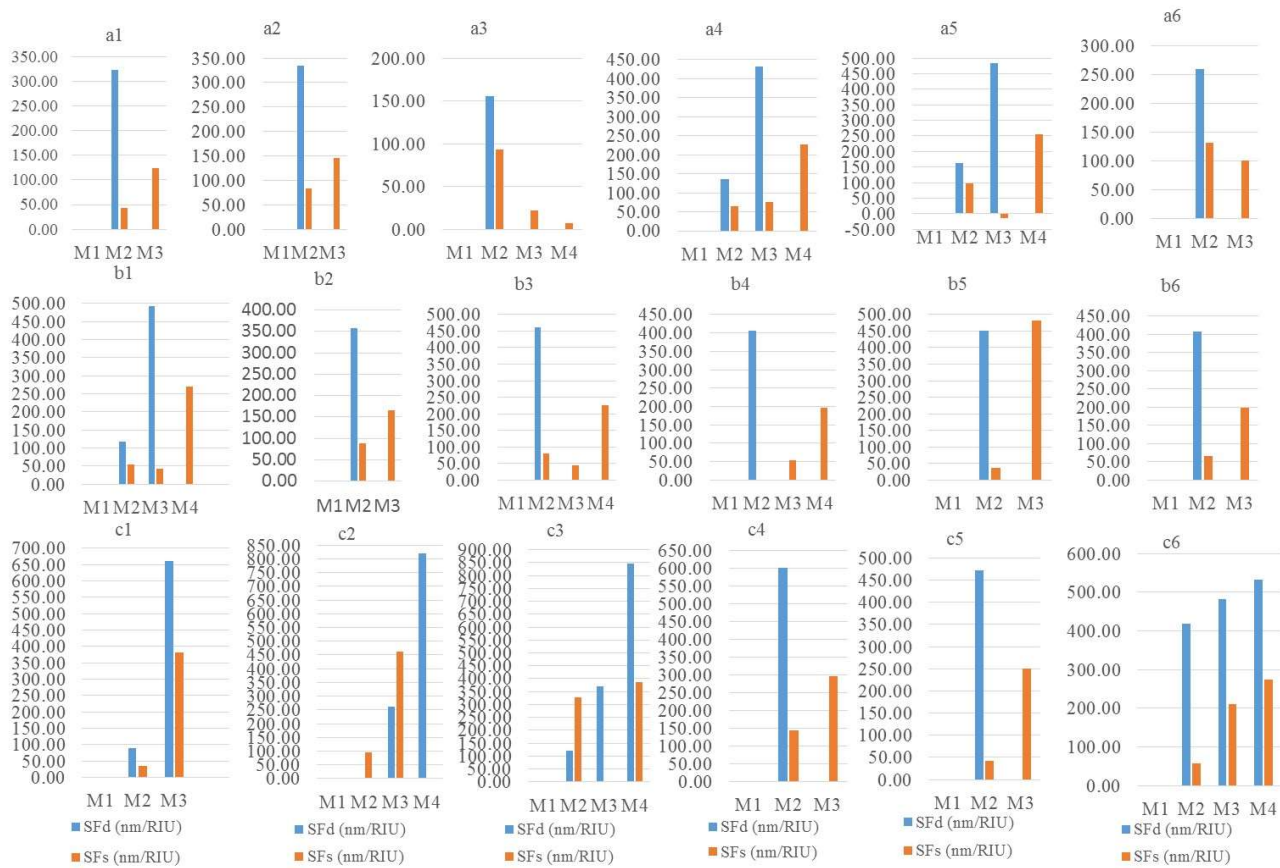

**Fig.6.** The sensitivity factor of all plasmonic modes for SNPs in dispersed and sensed mediums (SFd and SFs).